\documentclass[twocolumn,showpacs,preprintnumbers,amsmath,amssymb]{revtex4}
\usepackage{graphicx}
\usepackage{dcolumn}
\usepackage{bm}
\def\w6{{$\hat{W}_6$}}

\begin{document}

\title{Signature of Nearly Icosahedral Structures in Liquid and Supercooled Liquid Copper}

\author{P. Ganesh}
\author{M. Widom}
\affiliation{Carnegie Mellon University\\
Department of Physics\\
Pittsburgh, PA  15213}

\date{\today}

\begin{abstract}
A growing body of experiments display indirect evidence of icosahedral
structures in supercooled liquid metals.  Computer simulations provide
more direct evidence but generally rely on approximate interatomic
potentials of unproven accuracy.  We use first-principles molecular
dynamics simulations to generate realistic atomic configurations,
providing structural detail not directly available from experiment,
based on interatomic forces that are more reliable than conventional
simulations.  We analyze liquid copper, for which recent experimental
results are available for comparison, to quantify the degree of local
icosahedral and polytetrahedral order.
\end{abstract}


\maketitle

\section{\label{sec:intro}Introduction}

Turnbull~\cite{Turnbull1,Turnbull2,Turnbull3} established that
metallic liquids can be supercooled if heterogeneous nucleation can be
reduced or avoided.  Later, Frank hypothesized that the supercooling
of liquid metals might be due to frustrated packing of icosahedral
clusters. Icosahedral clustering of 12 atoms about a sphere is
energetically preferred to crystalline (e.g. FCC, HCP or BCC) packings
for the Lennard-Jones (L-J) pair potentials.  The icosahedron is
favorable because it is made up entirely of four-atom tetrahedra, the
densest-packed cluster possible.  Local icosahedral order cannot be
propagated throughout space without introducing defects.

Remarkably, the frustration of packing icosahedra is relieved in a
curved space, where a perfect 12-coordinated icosahedral packing
exists~\cite{SM82,Nelson83,Sethna83}.  Disclination line defects must
be introduced into this icosahedral crystal in order to ``flatten''
the structure and embed it in ordinary three dimensional space.  Owing
to the 5-fold rotational symmetry of an icosahedron, the disclination
lines are of type $72^{\circ}$.  The negative disclination line
defects that are needed to flatten the structure cause increased
coordination numbers of 14, 15 or 16.  Large atoms, if present, would
naturally assume high coordination number and aid in the formation of
a disclination line network.

Many studies of Lennard-Jones systems have tested Frank's hypothesis.
Hoare~\cite{Hoare} found that for clusters ranging between 2 to 64
atoms at least three types of ``polytetrahedral'' noncrystalline
structures exist, with a higher binding energy than HCP or FCC
structures with the same number of atoms.  Honeycutt and
Andersen~\cite{Honeycutt} found the crossover cluster size between
icosahedral and crystallographic ordering around a cluster size of
5000 atoms.  They also introduced a method to count the number of
tetrahedra surrounding an interatomic bond.  This number is 5 for
local icosahedral order.  Steinhardt, Nelson and
Ronchetti~\cite{Steinhardt} introduced the orientational order
parameter \w6 to demonstrate short range icosahedral
order.

Many other simulations have been performed on pure elemental metals
and metal alloys, using a modified Johnson potential~\cite{Egami},
embedded atom potentials~\cite{Chen-Liu,Sadigh}, the Sutton-Chen (SC)
many body potential~\cite{Goddard}, to name a few.  These potentials
model the interatomic interactions with varying, and generally
uncontrolled, degrees of accuracy.  Ab-initio studies on liquid
Copper~\cite{Vanderbilt,Hafner}, Aluminum~\cite{Valladares} and
Iron~\cite{Kresse} achieve a high degree of realism and accuracy, but
have not been analyzed from the perspective of icosahedral ordering.
Nevertheless, recent Ab-Initio studies on Ni and
Zr~\cite{Jakse_Ni,Jakse_Zr} have been done with this perspective and
find that with supercooling the degree of icosahedral ordering
increases in Ni while in Zr BCC is more favored.

X-ray diffraction measurements of electrostatically levitated droplets
of Ni~\cite{KeltonNi} found evidence of distorted icosahedral short
ranged order. Neutron scattering studies of deeply undercooled
metallic melts~\cite{Holland2} observed the characteristic shoulder on
the second peak of the structure factor, which has been identified as
a signature of icosahedral short range
order~\cite{NelsonWidom,Sachdev}.  The shoulder height increases with
decrease in temperature.

A recent XAS (X-ray Absorption Spectroscopy) experiment on liquid and
undercooled liquid Cu by Di Cicco {\em et al.}~\cite{Cicco} isolated
the higher order correlation functions. They applied Reverse
Monte-Carlo (RMC) refinement~\cite{Greevy1,Greevy2,Wang}
simultaneously to diffraction and XAS data to construct a model of the
disordered system compatible with their experimental data.  They
analyze the three body angular distribution function N($\theta$) and
also the orientational order parameter \w6.  Their conclusion was that
weak local icosahedral order could be observed in their sample.  This
experiment provided the most direct experimental evidence to-date of
the existence of icosahedra in a liquid metal.

Motivated by these results, we explore the structures of liquid and
undercooled liquid metals using first principles simulations.  First
principles calculations achieve the most realistic possible
structures, unhindered by the intrinsic inaccuracy of phenomenological
potentials, and with the ability to accurately capture the chemical
nature and distinctions between different elements and alloys.  We use
the VASP (Vienna Ab-initio Simulation Package~\cite{VASP,VASP2}) code
which solves the quantum mechanical interacting many-body problem
using electronic density functional theory.  These forces are
incorporated into a molecular dynamics simulation. The trade-off for
increased accuracy is a decrease in the system sizes we can study, so
we can only observe local order, not long range. Also we are limited
to short time scales.

Our analysis covers methods that have previously been fruitful. We
look at the radial distribution function, the structure factor, the
three body angular distribution function, which is simply related to
the three body correlation function, the \w6 parameter as discussed
above, and the Honeycutt and Andersen bond statistics method
~\cite{Honeycutt}.

The extent of the icosahedral order that we observe in simulation is
qualitatively in agreement with recent experiments~\cite{Cicco}.  At
high temperatures we found that structural properties of liquid Cu
strongly resembled a maximally random jammed~\cite{MRJ} hard sphere
configuration.  From this we conclude that a nearly universal
structure exists for single component systems whose energetics are
dominated by repulsive central forces. The degree of icosahedral order
is not great, presumably due to the frustration of icosahedra, but it
does show a tendency to increase as temperature drops.

Section~\ref{sec:FPM} describes our first principles molecular
dynamics method in greater detail. The next section,
section~\ref{sec:MD_Cu} discusses our study on copper. Here we
introduce the radial distribution function $g(r)$, the liquid
structure factor $S(q)$, the \w6 bond orientational order parameter,
the three body angular distribution function $N(\theta)$, and the
Honeycutt and Andersen analysis method. We conclude, in
section~\ref{sec:conclude}, with some thoughts about enhancing
icosahedral order by alloying with a fraction of smaller and larger
atoms.

\section{\label{sec:FPM}First Principles Method}

First principles simulation is an incisive, powerful and
well-developed tool based on a quantum mechanical treatment of the
electrons responsible for interatomic bonding. Since the method is
based on fundamental physical laws and properties of atoms, it can be
applied to a wide variety of metals, including alloys, and yields the
energy and forces computationally without any adjustable free
parameters.

Our $\mathit{ab-initio}$ molecular dynamics simulation program,
VASP~\cite{VASP,VASP2}, solves the N-body quantum mechanical
interacting electron problem using electronic density functional
theory, under the Generalized Gradient Approximation (GGA). We use the
projector-augmented wave~\cite{PAW,KJ_PAW} (PAW) potentials as
provided with VASP. Calculation times grow nearly as the third power
of the number of atoms, limiting our studies to sample sizes of around
a hundred atoms.

In first-principles molecular dynamics, although interatomic forces
and energies are calculated quantum mechanically, we still treat the
atomic motions classically, using the Born-Oppenheimer approximation.
We use Nose dynamics~\cite{Nose} to simulate in the canonical ensemble
at fixed mean temperature. The system was well equilibrated before
data was considered for analysis. The simulation started with a random
configuration, at a temperature high enough to ensure a liquid state,
and was allowed to equilibrate at this high temperature.
Subsequently, lower temperatures were simulated starting from previous
configurations. All calculations were $\Gamma$ point calculations (a
single `k' point).

We took N=100 Cu atoms and applied periodic boundary conditions in an
orthorhombic cell. Our unequal lattice parameters avoid imposing a
characteristic length on the system. The simulations were done at
three different temperatures, T=1623K, 1398K and 1313K in order to
compare with Di Cicco's experiments~\cite{Cicco}. The melting point of
copper is T=1356K, so samples at 1623K and 1398K are in the liquid
regime, while the one at 1313K is undercooled. We used number
densities of 0.0740~\AA$^{-3}$, 0.0758~\AA$^{-3}$ and
0.0764~\AA$^{-3}$ respectively at T=1623K, 1398K and 1313K. These were
obtained from a fit of the XRD experimental volume per
particle~\cite{Waseda} to a straight line versus temperature.
Starting from configurations that had been previously equilibrated at
slightly different densities, transients of about 250 steps (1fs per
step) passed prior to the onset of equilibrium fluctuations of the
energy. After the transient, a total of 3000 MD steps were taken at
each temperature, for a total simulation time of 3ps. The run time was
around 480hrs on a 2.8 GHz Intel Xeon processor for each
temperature. Subsequently, two configurations from the highest
temperature run, widely separated in time, were selected and used as
the starting configuration for two runs at T=1398K, with proper
scaling of densities. After an initial transient of 250 steps, these
runs were further continued at T=1313K as well as at T=1398K. The four
runs, two at T=1398K and two at T=1313K were carried on for 1ps,
during which time the energy showed equilibrium fluctuations. All of
these runs have been used for analyzing the local order in liquid
copper.

For a few selected configurations, a conjugate-gradient algorithm was
used to relax the ions to their instantaneous ground state, to explore
their inherent structures~\cite{Stillinger}. Surprisingly all of our
samples partially crystallized, based on visual observation. Further
efforts to obtain quenched amorphous structures used steepest-descent
minimization and molecular-dynamics with a linear temperature ramp,
followed by steepest-descent minimization. Again the structures
partially crystallized. There appear to be no fully amorphous relaxed
structures accessible for Cu using these methods.

\section{\label{sec:MD_Cu}Results }
 
\subsection{\label{sec:gofr_Cu}Radial Distribution Function $g(r)$}

\begin{figure}
\includegraphics[width=3in,angle=-90]{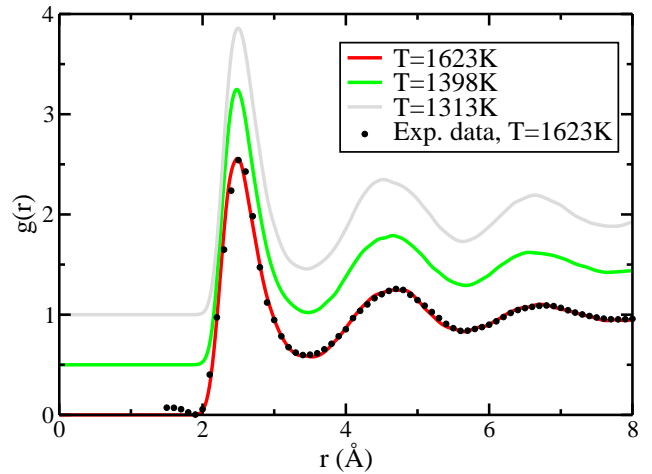}
\caption{\label{fig:Cugr}(Color online) Simulated liquid Cu radial
distribution function, $g(r)$, at three different temperatures. The
simulated curve at T=1623K matches well with the experimental XRD
(X-Ray Diffraction) result~\cite{Waseda} interpolated to T=1623K. (The
simulated $g(r)$ at T=1313K and T=1398K have been shifted up for
visual clarity)}
\end{figure}

The radial distribution function, $g(r)$, is proportional to the
density of atoms at a distance $r$ from another atom. We calculate
$g(r)$ by forming a histogram of bond lengths. We use the repeated
image method to obtain bond lengths greater than half the box size,
and anticipate $g(r)$ in this range may be strongly influenced by
finite size effects. We then smooth out with a gaussian of standard
deviation 0.05~\AA. Fig.~\ref{fig:Cugr} shows the $g(r)$ we obtained
at the three different temperatures. Our $g(r)$ at T=1623K, compares
well with $g(r)$ interpolated from XRD experiments~\cite{Waseda}, with
the two curves overlapping almost everywhere except for a small
disagreement in the position of the first peak. Results from neutron
diffraction experiment~\cite{Eder} at T=1393K, compare well with our
$g(r)$ at T=1398K. Comparisons with the $g(r)$ for Cu at 1500K from
the $ab-initio$ MD studies by Hafner {\em et. al.}~\cite{Hafner} and
Vanderbilt {\em et. al.}~\cite{Vanderbilt} finds that the heights of
their first peak match well with our $g(r)$ (interpolated to
T=1500K). But their peak positions are shifted slightly to the left of
ours (ours is at 2.50~\AA). The $g(r)$ from an embedded-atom method
(EAM) model for Cu~\cite{Sadigh} which matches almost exactly with the
XRD data at T=1773K, is also consistent with our extrapolated $g(r)$
at this temperature.

The growth in height of the peaks in the supercooled system at T=1313K
suggests an increase of some type of order. However this order is not
related to the crystalline FCC equilibrium phase, as we show in the
following subsections.

\begin{figure}
\includegraphics[width=3in,angle=-90]{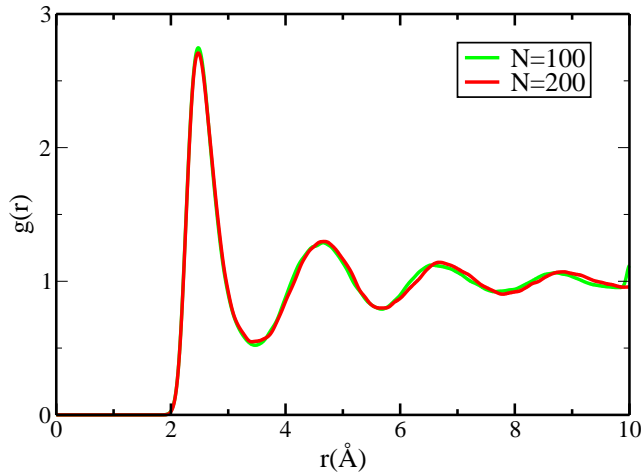}
\caption{\label{fig:gr_N} (Color online) Simulated liquid Cu radial
distribution function, $g(r)$, at T=1398K for N=100 and N=200 atoms. A
weak finite size effect is observed after about $r$=5\AA.}
\end{figure}

To test for finite size effects in our N=100 atom system, we ran a
separate simulation for N=200 atoms at the intermediate temperature
T=1398K (Fig.~\ref{fig:gr_N}). The first and second peaks of $g(r)$
for both the system sizes compare very well.  There is a small but
significant difference in the depth of the first minimum, then
systematic differences between the curves beyond 5\AA.  From this we
conclude that the finite size effect is not important at small $r$
values, but for larger values of $r$ (beyond 5~\AA) there is a weak
finite size effect. The three body angular distributions, and the \w6
histograms of the N=200 and the N=100 runs, are also comparable,
suggesting that N=100 is sufficient for studies of local order of the
types we consider here.

We calculate the coordination number from the radial distribution
function $g(r)$. We choose a cutoff distance near the first minimum of
$g(r)$, at $R_{cut}$=3.4~\AA. The precise location of the minimum is
difficult, and its variation with temperature is smaller than the
error in locating its position (Fig.~\ref{fig:Cugr}), so that we don't
change the value of $R_{cut}$ with temperature. With this value of
$R_{cut}$ we find an average coordination number ($N_c$) of 12.3 which
is nearly independent of temperature ($N_c$ changes from 12.1 at high
temperature to 12.5 with supercooling). However, our range of
evolution of $N_c$ is small compared to the case of
Ni~\cite{Jakse_Ni}, since our degree of supercooling is much lower
(3\%) than theirs (17\%). We are not able to achieve a higher degree
of supercooling of Cu as mentioned earlier in the paper.

\subsection{\label{sec:Sq_Cu}Liquid Structure Factor S(q)}

 The liquid structure factor $S(q)$ is related to the radial
 distribution function $g(r)$ of a liquid with density $\rho$ by,
\begin{equation}
\label{eq:Sqeq}
S(q)=1+4 \pi \rho \int \limits_{0}^{\infty}[g(r)-1] {\sin(qr) \over qr} r^2 dr. 
\end{equation}
Evidently, one needs a knowledge of the radial distribution function
up to large values of $r$ to get a good $S(q)$. In our first
principles simulation, we are restricted to small values of $r$, due
to our small system sizes, so we need a method to get $S(q)$ from our
limited range $g(r)$ function.

Baxter developed a method~\cite{Baxter,Jolly} to extend $g(r)$ beyond
the size of the simulation cell. The method exploits the short ranged
nature of the direct correlation function $c(r)$, which has a range
similar to the interatomic interactions~\cite{Ashcroft}, as opposed to
the $g(r)$ which is much long ranged. The exact relation that connects
these two functions is the Ornstein-Zernike relation,
\begin{equation}
\label{eq:O-Z}
h(r)=c(r)+\rho \int h(|{\bf r}-{\bf r'}|)c(|{\bf r'}|)d{\bf r'}
\end{equation}
where $h(r)=g(r)-1$.
  
Assuming that $c(r)$ vanishes beyond a certain cutoff distance $r_c$,
Baxter obtained a pair of equations, valid for $r<r_c$.
The remarkable property of this method is that if we know $h(r)$ over
a range $0\le r \le r_c$, then we can obtain $c(r)$ over its entire
range (from 0 to $r_c$), which implicitly determines $h(r)$ over its
entire range (from 0 to $\infty$) through Eq.~(\ref{eq:O-Z}).
 
 We solve the Baxter's equations iteratively to obtain the full direct
 correlation function.  A complete knowledge of the direct correlation
 function gives us the structure factor $S(q)$ in terms of its fourier
 transform $\hat{c}(q)$,
\begin{equation}
S(q)={1\over 1-\rho \hat c(q)}
\label{OZink}
\end{equation}
where,
\begin{equation}
\label{eq:cq}
\hat c(q)=4 \pi \int \limits_{0}^{\infty}r^2 c(r) {\sin(qr)\over qr}dr. 
\end{equation}
The $S(q)$ showed good convergence with different choices of $r_c$,
and a choice of $r_c$=5\AA~ seemed appropriate because it was one half
of our smallest simulation cell edge length. Even though in metals
there are long range oscillatory Friedel oscillations, our ability to
truncate $c(r)$ at $r_c$=5\AA, shows that these are weak compared with
short range interactions.

\begin{figure}
\includegraphics[width=3in,angle=-90]{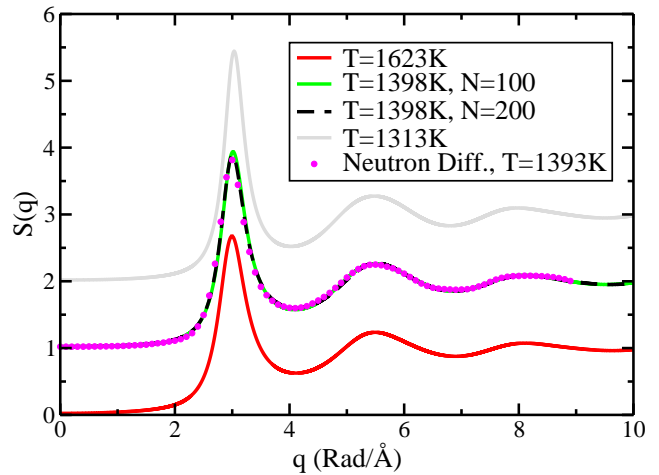}
\caption{\label{fig:Cusq} (Color online) Liquid structure factor
  $S(q)$ as obtained from the simulated radial distribution function
  $g(r)$ at T=1398K compared with the $S(q)$ from neutron diffraction
  at T=1393K~\cite{Eder}. The calculated $S(q)$ at the other two
  temperatures are also plotted, and show the expected temperature
  behavior. (The $S(q)$ at T=1313K and T=1398K have been shifted up
  for visual clarity) }
\end{figure}

 Fig.~(\ref{fig:Cusq}) compares the calculated $S(q)$ at our three
different temperatures, and the experimental neutron $S(q)$ at
T=1393K~\cite{Eder}. The calculated $S(q)$ at T=1398K compares well
with the experiment at all values of $q$. The $S(q)$ from the larger
system is in better agreement with the experiment. Comparison between
the two system sizes suggest again that the finite size effects are
significant but not important, and N=100 is good enough to get a
representative liquid structure. No resolution correction was applied
to the experimental data, and moreover it was smoothed. Both of these
cause a decrease in the height of the actual $S(q)$, which becomes
quite appreciable at the first peak. As a test, we also applied a
resolution correction to our simulated $S(q)$ (not shown), which
reduced the height of the first peak bringing it in closer agreement
with the experimental value. Nevertheless the overall excellent
agreement shows that the first principles simulation with only N=100
atoms is able to produce representative structures at T=1398K. This
enables us to make further studies of the local icosahedral and
polytetrahedral order in liquid and supercooled liquid copper.

As mentioned earlier in the introduction, one signature of icosahedral
short range order is the splitting of the second peak of
$S(q)$~\cite{NelsonWidom,Sachdev}. Even though we observe weak
icosahedral order in Cu (discussed later in this paper), we do not
observe a clear splitting of the second peak in $S(q)$
(Fig.~\ref{fig:Cusq}), but we do observe a broadening as we lower the
temperature. We think that the absence of splitting could be because
of our low degree of supercooling.

\subsection{\label{sec:W6_Cu} Bond Orientation Order Parameters}
  
\begin{figure}
\includegraphics[width=3in,angle=-90]{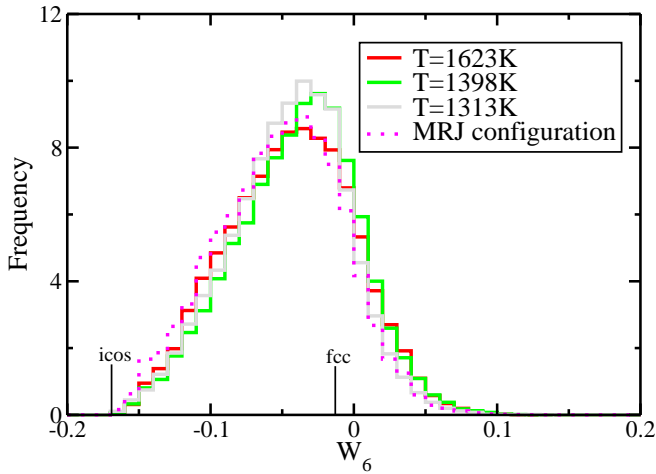}
\caption{\label{fig:Cuw6} (Color online) Simulated \w6 distributions
for liquid Cu. Ideal icosahedron and FCC values are indicated.}
\end{figure}

As introduced by Steinhardt, {\em et al.}~\cite{Steinhardt}, the
$\hat{W}_l$ parameters are a measure of the local orientational order
in liquids and undercooled liquids. To calculate $\hat{W}_l$, the
orientations of bonds from an atom to its neighboring atoms are
projected onto a basis of spherical harmonics. Rotationally invariant
combinations of coefficients in the spherical harmonics expansion are
then averaged over many atoms in an ensemble of configurations. The
resulting measures of local orientational order can be used as order
parameters to characterize the liquid structures. For an ideal
icosahedral cluster, $l=6$ is the minimum value of $l$ for which \w6
$\ne 0$. Table~\ref{tab:W6} enumerates \w6 values for different ideal
clusters. We see that the ideal icosahedral value of \w6 is far from
other clusters, making it a good icosahedral order indicator.

We choose the cutoff distance to specify near neighbors at
$R_{cut}$=3.4~\AA~ as before. Our value of $R_{cut}$ is significantly
greater than that of Di Cicco, {\em et al.}. Our \w6 distributions
(Fig.~\ref{fig:Cuw6}) show strong asymmetry favoring negative values
with tails extending towards the ideal icosahedron value. Because the
histogram vanishes as \w6 approaches its limiting negative value we
see that there are essentially no perfectly symmetric undistorted
icosahedra present in our simulation.  However, a significant fraction
do have \w6 values close to the icosahedral value.

A \w6 analysis was performed for a $10^4$ atom maximally random jammed
hard sphere configuration~\cite{MRJ}. The diameter of the hard spheres
was rescaled so that the position of the main peak of the resulting
$g(r)$ matched the value $r=2.5$\AA~ found for Cu at T=1623K.  The
$R_{cut}$ value for the MRJ configuration was taken near the first
minimum of the $g(r)$ at $r=3.3$\AA. Remarkably, the \w6 distribution
of the MRJ configuration (Fig.~\ref{fig:Cuw6}) is similar to the
distribution for liquid Cu at high temperature, suggesting that the
structure of Cu under this condition is dominated by strongly
repulsive short-range central forces.

As we lower the temperature of liquid Cu, the mean value of \w6 drops
and the peak of the \w6 distribution shifts to the left.  However, the
peak never moves below \w6 = -0.05, and the tail of the distribution
at negative \w6 shows no strong temperature dependence.  It seems that
there is no change in the number of nearly icosahedral clusters as the
temperature drops into the supercooled regime, possibly a result of
the frustration of icosahedral packing.  Our liquid has a single
component, so there is no natural way to introduce disclinations. This
inhibits the growth of a population of atoms with \w6 close to its
ideal icosahedral value.

Comparing our result with that of Di Cicco {\em et al.} at T=1313K, we
see that our curve is more asymmetric towards negative values than Di
Cicco's, so that we see a greater fraction of atoms near the ideal
icosahedral value of \w6. The discrepancy probably lies in the
difference between the two methods used to generate the positional
configurations (the difference is even greater if we use Di Cicco's
value of $R_{cut}$). Their configurations were obtained using Reverse
Monte Carlo (RMC), which does not guarantee accurate configurations.
Our first principles method should be more accurate in determining
these configurations. Of course, Di Cicco's configurations {\em are}
consistent with experimentally measured three-body correlations. It
would be of great interest to see if our configurations are also
consistent with the raw experimental data.  The differences in \w6
distributions should not be overstated - the experiment and our
simulations both show that liquid and supercooled liquid copper has
weak but non-negligible icosahedral order.

\begin{table}
\caption{\label{tab:W6} $\hat{W}_6$ values for a few clusters}
\begin{tabular}{|r||r|r|r|r|}
\hline

Cluster & HCP & FCC & ICOS & BCC \\
\hline
No. of atoms & 12 & 12 & 12 & 14 \\
\hline 
$\hat{W}_6$ &  -0.012 & -0.013 & -0.169 & +0.013 \\
\hline
\end{tabular}
\end{table}

\subsection{\label{sec:Bang}Bond Angle Distribution N($\theta$):}

\begin{figure}
\includegraphics[width=3in,angle=-90]{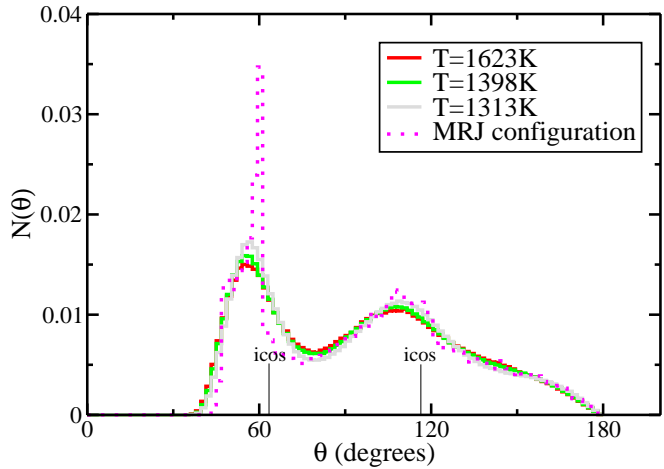}
\caption{\label{fig:Cutheta} (Color online) Distribution of N($\theta$) for liquid
  Cu. Ideal icosahedron values are indicated.}
\end{figure}

The bond angle distribution N($\theta$) is a simple type of three-body
correlation function. Let $\theta$ be the angle between bonds from a
single atom to two neighbors, and define N($\theta$) as the
probability density for angle $\theta$, normalized such that the total
probability, $\int{N(\theta)d\theta}$=1. The distribution for the
central atom of an ideal 13-atom icosahedral cluster, shows peaks at
63.4$^{\circ}$, 116.4$^{\circ}$ and 180.0$^{\circ}$. For other
crystallographic clusters, like HCP, FCC, and BCC, we see peaks at
60$^{\circ}$, 90$^{\circ}$ and 120$^{\circ}$. Angles around
60$^{\circ}$ degrees indicate nearly equilateral triangles that may
well belong to tetrahedra.

Fig.~\ref{fig:Cutheta} shows the distributions for copper at three
temperatures. We have chosen the same value of $R_{cut}$ that was used
to obtain \w6 in section~\ref{sec:W6_Cu} . The distribution function
shows maxima at 56$^{\circ}$ and 110$^{\circ}$ with a minimum around
80$^{\circ}$. Our result is similar to that of Di Cicco (they show
only T=1313K), but with more pronounced minimum and second
maximum. The peak around 60$^{\circ}$ shows an abundance of nearly
equilateral triangles, indicating the presence of tetrahedrons, which
can pack to form icosahedra. The minimum close to 90$^{\circ}$ shows
that there aren't many cubic clusters. We also see that the high-angle
tail at high temperature turns into a broad maximum at low temperature
centered around 165$^{\circ}$. This may represent a shifting of the
ideal 180$^{\circ}$ peak caused by cluster distortion. The ordering
increases as temperature decreases, indicating that the number of
nearly equilateral triangles increases when the liquid is undercooled,
probably caused by an increase in polytetrahedral order with
undercooling.

  The distribution of the MRJ configurations (same $R_{cut}$ as
  defined in section~\ref{sec:W6_Cu}) shows a sharp peak at exactly
  60$^{\circ}$, a broad peak at 110$^{\circ}$ and a minimum around
  90$^{\circ}$.  The peak at 60$^{\circ}$ shows an overwhelming
  presence of perfectly equilateral triangular faces, which are easily
  formed when 3 hard spheres come in contact with each other. But the
  minimum around 90$^{\circ}$ and a second maximum nearer to
  110$^{\circ}$ as opposed to 120$^{\circ}$, suggests that the local
  order is not FCC or HCP.  This feature of the MRJ configuration
  agrees qualitatively with the angular distribution of liquids,
  implying an underlying universal structure for systems whose
  energetics are dominated by repulsive central forces. But the
  quantitative differences also emphasize the necessity to exactly
  model an atomic liquid to study its local environments, and quantify
  polytetrahedral order.

\subsection{\label{sec:HA}Honeycutt and Andersen analysis}

\begin{figure}
\includegraphics[width=3in,angle=-90]{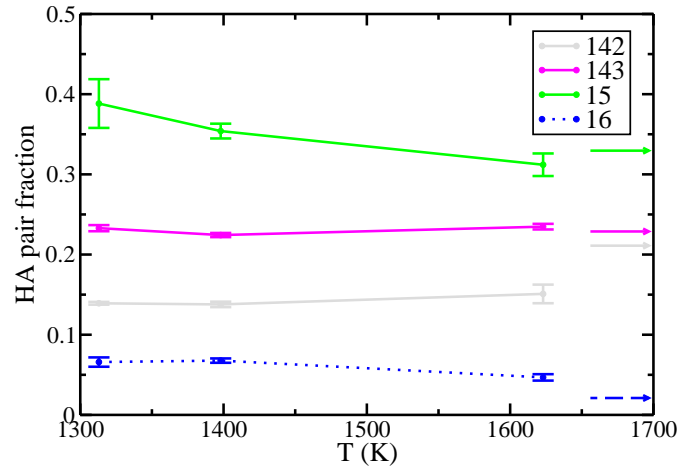}
\caption{\label{fig:HA} (Color online) Honeycutt-Andersen pair
  fractions for 142's (FCC and HCP forming +72$^{\circ}$
  disclination), 15's (icosahedron) and 16's ($-72^{\circ}$
  disclination) at different temperatures. The increase in 15's as
  temperature is reduced show increased icosahedral ordering with
  supercooling. The corresponding HA values for the MRJ configurations are
  indicated on the right side of the plot.}
\end{figure}

Honeycutt and Andersen~\cite{Honeycutt} introduced a useful assessment
of local structure surrounding interatomic bonds.  We employ a
simplified form of their analysis, counting the number of common
neighbors shared by a pair of near-neighbor atoms.  This identifies
the number of atoms surrounding the near-neighbor bond and usually
equals the number of edge-sharing tetrahedra whose common edge is the
near-neighbor bond. We assign a set of three indices to each bond. The
first index is 1 if the root pairs are bonded (separation less than or
equal to $R_{cut}$).  The second index is the number of near-neighbor
atoms common to the root pairs, and the third index gives the number
of near-neighbor bonds between these common neighbors. We take
$R_{cut}$=3.4\AA~ as before. Note that the Honeycutt and Andersen
fractions depend sensitively on $R_{cut}$, making precise quantitative
comparisons with other prior studies difficult.

In general, 142's are characteristic of close packed structures (FCC
and HCP) and 143's are characteristic of distorted
icosahedras~\cite{Ni-Ag}.  They can also be considered as
+72$^{\circ}$ disclinations~\cite{SM82,Nelson83,Sethna83}.  Likewise,
15's are characteristic of icosahedra, and 16's indicate -72$^{\circ}$
disclinations.  Fig.~\ref{fig:HA} shows the 14's, 15's and the 16's
for liquid Cu at the three temperatures. The error bars shown were
calculated by breaking the data into three subsets. The 14's have been
separated into 142's and 143's.  The remaining 14's are mostly 144's
with fraction around 0.04. The fraction of 142's and 143's holds
steady with temperature, with the icosahedral fraction always
exceeding the close packed fraction. As the temperature drops, the
fraction of 15's grows. At each of the three temperatures, the 15's
are mainly comprised of 154's ( characteristic of distorted
icosahedra) and 155's (characteristic of perfect icosahedra), with the
154's slightly higher than the 155's. Of all the 16's, the 166's are
the highest and steadily increase with lowering of temperature. The
166's indicate the -72$^{\circ}$ disclination lines, which relieve the
frustration of icosahedral order.

 These trends indicate a weak increase in polytetrahedral ordering
with supercooling.  The same trend was observed in simulations based
on Sutton-Chen potentials~\cite{Goddard} except for the fact that our
142's are slightly higher compared to their 142's.

  For comparison, the values for a maximally random jammed
packing~\cite{MRJ} ($R_{cut}$=3.3\AA) are shown in Fig.~\ref{fig:HA}.
These values are fairly close to liquid Cu at high temperature, and
also to a similar common neighbor analysis of dense random-packing of
hard spheres~\cite{Clarke}. These results are consistent with our
previous observation for the \w6 distribution and N($\theta$), that a
nearly universal structure arises at high temperature, dominated by
repulsive central forces.

\section{\label{sec:conclude} Conclusion}

This study quantifies icosahedral and polytetrahedral order in
supercooled liquid copper. While the structural properties of high
temperature liquid Cu are close to a maximally random jammed
structure~\cite{MRJ}, proper modeling of atomic interactions is
essential to capture the behavior of an element at liquid and
supercooled temperatures. A first-principles simulation is the most
reliable means of achieving this. We find small but significant
disagreement with analysis based on Reverse Monte-Carlo simulation.

Supercooled liquid copper shows a slight increase in icosahedral and
polytetrahedral order as temperature drops, which is consistent with
recent experiments~\cite{KeltonNi,Holland2,Cicco}. The frustration of
icosahedrons in the one component liquid inhibits formation of perfect
icosahedra, giving rise to defective icosahedrons. Alloying with
larger atoms might relieve the frustration of packing icosahedrons by
encouraging the formation of -72$^{\circ}$
disclinations~\cite{Jakse_alloy1,Jakse_alloy2}. Alloying with smaller
atoms can relieve frustration of individual icosahedrons by placing
the smaller atom at the center~\cite{MaNature}. Alloying with larger
and smaller atoms simultaneously thus offers the chance to optimize
icosahedral order. Work is in progress in achieving the
same~\cite{Fe_alloy}.

\begin{acknowledgements}
This work was supported in part by NSF/DMR-0111198.  We thank Sal
Torquato for providing maximally random jammed configurations of
hard-spheres.
\end{acknowledgements}

\bibliography{ico_cu}

\begin{thebibliography}{45}
\expandafter\ifx\csname natexlab\endcsname\relax\def\natexlab#1{#1}\fi
\expandafter\ifx\csname bibnamefont\endcsname\relax
  \def\bibnamefont#1{#1}\fi
\expandafter\ifx\csname bibfnamefont\endcsname\relax
  \def\bibfnamefont#1{#1}\fi
\expandafter\ifx\csname citenamefont\endcsname\relax
  \def\citenamefont#1{#1}\fi
\expandafter\ifx\csname url\endcsname\relax
  \def\url#1{\texttt{#1}}\fi
\expandafter\ifx\csname urlprefix\endcsname\relax\def\urlprefix{URL }\fi
\providecommand{\bibinfo}[2]{#2}
\providecommand{\eprint}[2][]{\url{#2}}

\bibitem[{\citenamefont{Turnbull}(1950{\natexlab{a}})}]{Turnbull1}
\bibinfo{author}{\bibfnamefont{D.}~\bibnamefont{Turnbull}},
  \bibinfo{journal}{J. Appl. Phys.} \textbf{\bibinfo{volume}{21}},
  \bibinfo{pages}{1022} (\bibinfo{year}{1950}{\natexlab{a}}).

\bibitem[{\citenamefont{Turnbull}(1950{\natexlab{b}})}]{Turnbull2}
\bibinfo{author}{\bibfnamefont{D.}~\bibnamefont{Turnbull}},
  \bibinfo{journal}{J. Metals} \textbf{\bibinfo{volume}{188}},
  \bibinfo{pages}{1144} (\bibinfo{year}{1950}{\natexlab{b}}).

\bibitem[{\citenamefont{Turnbull and Cech}(1950)}]{Turnbull3}
\bibinfo{author}{\bibfnamefont{D.}~\bibnamefont{Turnbull}} \bibnamefont{and}
  \bibinfo{author}{\bibfnamefont{R.~E.} \bibnamefont{Cech}},
  \bibinfo{journal}{J. Appl. Phys.} \textbf{\bibinfo{volume}{21}},
  \bibinfo{pages}{804} (\bibinfo{year}{1950}).

\bibitem[{\citenamefont{Sadoc and Mosseri}(1982)}]{SM82}
\bibinfo{author}{\bibfnamefont{J.~F.} \bibnamefont{Sadoc}} \bibnamefont{and}
  \bibinfo{author}{\bibfnamefont{R.}~\bibnamefont{Mosseri}},
  \bibinfo{journal}{Phil. Mag. B} \textbf{\bibinfo{volume}{45}},
  \bibinfo{pages}{467} (\bibinfo{year}{1982}).

\bibitem[{\citenamefont{Nelson}(1983)}]{Nelson83}
\bibinfo{author}{\bibfnamefont{D.~R.} \bibnamefont{Nelson}},
  \bibinfo{journal}{Phys. Rev. Lett.} \textbf{\bibinfo{volume}{50}},
  \bibinfo{pages}{982} (\bibinfo{year}{1983}).

\bibitem[{\citenamefont{Sethna}(1983)}]{Sethna83}
\bibinfo{author}{\bibfnamefont{J.~P.} \bibnamefont{Sethna}},
  \bibinfo{journal}{Phys. Rev. Lett.} \textbf{\bibinfo{volume}{51}},
  \bibinfo{pages}{2198} (\bibinfo{year}{1983}).

\bibitem[{\citenamefont{Hoare}(1972)}]{Hoare}
\bibinfo{author}{\bibfnamefont{M.~R.} \bibnamefont{Hoare}},
  \bibinfo{journal}{J. Cryst. Growth} \textbf{\bibinfo{volume}{17}},
  \bibinfo{pages}{77} (\bibinfo{year}{1972}).

\bibitem[{\citenamefont{Honeycutt and Andersen}(1987)}]{Honeycutt}
\bibinfo{author}{\bibfnamefont{J.~D.} \bibnamefont{Honeycutt}}
  \bibnamefont{and} \bibinfo{author}{\bibfnamefont{H.~C.}
  \bibnamefont{Andersen}}, \bibinfo{journal}{J. Phys. Chem.}
  \textbf{\bibinfo{volume}{91}}, \bibinfo{pages}{4950} (\bibinfo{year}{1987}).

\bibitem[{\citenamefont{Steinhardt et~al.}(1983)\citenamefont{Steinhardt,
  Nelson, and Ronchetti}}]{Steinhardt}
\bibinfo{author}{\bibfnamefont{P.~J.} \bibnamefont{Steinhardt}},
  \bibinfo{author}{\bibfnamefont{D.~R.} \bibnamefont{Nelson}},
  \bibnamefont{and}
  \bibinfo{author}{\bibfnamefont{M.}~\bibnamefont{Ronchetti}},
  \bibinfo{journal}{Phys. Rev. B} \textbf{\bibinfo{volume}{28}},
  \bibinfo{pages}{784} (\bibinfo{year}{1983}).

\bibitem[{\citenamefont{Tomida and Egami}(1995)}]{Egami}
\bibinfo{author}{\bibfnamefont{T.}~\bibnamefont{Tomida}} \bibnamefont{and}
  \bibinfo{author}{\bibfnamefont{T.}~\bibnamefont{Egami}},
  \bibinfo{journal}{Phys. Rev. B} \textbf{\bibinfo{volume}{52}},
  \bibinfo{pages}{3290} (\bibinfo{year}{1995}).

\bibitem[{\citenamefont{Kuiying et~al.}(1995)\citenamefont{Kuiying, Hongbo,
  Xiaoping, Quiyong, and Zhuangqi}}]{Chen-Liu}
\bibinfo{author}{\bibfnamefont{C.}~\bibnamefont{Kuiying}},
  \bibinfo{author}{\bibfnamefont{L.}~\bibnamefont{Hongbo}},
  \bibinfo{author}{\bibfnamefont{L.}~\bibnamefont{Xiaoping}},
  \bibinfo{author}{\bibfnamefont{H.}~\bibnamefont{Quiyong}}, \bibnamefont{and}
  \bibinfo{author}{\bibfnamefont{H.}~\bibnamefont{Zhuangqi}},
  \bibinfo{journal}{J. Phys. Condens. Matter} \textbf{\bibinfo{volume}{7}},
  \bibinfo{pages}{2379} (\bibinfo{year}{1995}).

\bibitem[{\citenamefont{Sadigh and Grimvall}(1996)}]{Sadigh}
\bibinfo{author}{\bibfnamefont{B.}~\bibnamefont{Sadigh}} \bibnamefont{and}
  \bibinfo{author}{\bibfnamefont{G.}~\bibnamefont{Grimvall}},
  \bibinfo{journal}{Phys. Rev. B} \textbf{\bibinfo{volume}{54}},
  \bibinfo{pages}{15742} (\bibinfo{year}{1996}).

\bibitem[{\citenamefont{Lee et~al.}(2003)\citenamefont{Lee, Cagin, Johnson, and
  Goddard}}]{Goddard}
\bibinfo{author}{\bibfnamefont{H.~J.} \bibnamefont{Lee}},
  \bibinfo{author}{\bibfnamefont{T.}~\bibnamefont{Cagin}},
  \bibinfo{author}{\bibfnamefont{W.~L.} \bibnamefont{Johnson}},
  \bibnamefont{and} \bibinfo{author}{\bibfnamefont{W.~A.}
  \bibnamefont{Goddard}}, \bibinfo{journal}{J. Chem. Phys.}
  \textbf{\bibinfo{volume}{119}}, \bibinfo{pages}{9858} (\bibinfo{year}{2003}).

\bibitem[{\citenamefont{Pasquarello et~al.}(1992)\citenamefont{Pasquarello,
  Laasonen, Car, Lee, and Vanderbilt}}]{Vanderbilt}
\bibinfo{author}{\bibfnamefont{A.}~\bibnamefont{Pasquarello}},
  \bibinfo{author}{\bibfnamefont{K.}~\bibnamefont{Laasonen}},
  \bibinfo{author}{\bibfnamefont{R.}~\bibnamefont{Car}},
  \bibinfo{author}{\bibfnamefont{C.}~\bibnamefont{Lee}}, \bibnamefont{and}
  \bibinfo{author}{\bibfnamefont{D.}~\bibnamefont{Vanderbilt}},
  \bibinfo{journal}{Phys. Rev. Letts.} \textbf{\bibinfo{volume}{69}},
  \bibinfo{pages}{1982} (\bibinfo{year}{1992}).

\bibitem[{\citenamefont{Kresse and Hafner}(1993{\natexlab{a}})}]{Hafner}
\bibinfo{author}{\bibfnamefont{G.}~\bibnamefont{Kresse}} \bibnamefont{and}
  \bibinfo{author}{\bibfnamefont{J.}~\bibnamefont{Hafner}},
  \bibinfo{journal}{Phys. Rev. B} \textbf{\bibinfo{volume}{48}},
  \bibinfo{pages}{13115} (\bibinfo{year}{1993}{\natexlab{a}}).

\bibitem[{\citenamefont{Valladares}(2004)}]{Valladares}
\bibinfo{author}{\bibfnamefont{A.~A.} \bibnamefont{Valladares}},
  \bibinfo{journal}{preprint}  (\bibinfo{year}{2004}).

\bibitem[{\citenamefont{Alfe et~al.}(2000)\citenamefont{Alfe, Kresse, and
  Gillan}}]{Kresse}
\bibinfo{author}{\bibfnamefont{D.}~\bibnamefont{Alfe}},
  \bibinfo{author}{\bibfnamefont{G.}~\bibnamefont{Kresse}}, \bibnamefont{and}
  \bibinfo{author}{\bibfnamefont{M.~J.} \bibnamefont{Gillan}},
  \bibinfo{journal}{Phys. Rev. B} \textbf{\bibinfo{volume}{61}},
  \bibinfo{pages}{132} (\bibinfo{year}{2000}).

\bibitem[{\citenamefont{Jakse and Pasturel}(2004)}]{Jakse_Ni}
\bibinfo{author}{\bibfnamefont{N.}~\bibnamefont{Jakse}} \bibnamefont{and}
  \bibinfo{author}{\bibfnamefont{A.}~\bibnamefont{Pasturel}},
  \bibinfo{journal}{J. Chem. Phys.} \textbf{\bibinfo{volume}{120}},
  \bibinfo{pages}{6124} (\bibinfo{year}{2004}).

\bibitem[{\citenamefont{Jakse and Pasturel}(2003)}]{Jakse_Zr}
\bibinfo{author}{\bibfnamefont{N.}~\bibnamefont{Jakse}} \bibnamefont{and}
  \bibinfo{author}{\bibfnamefont{A.}~\bibnamefont{Pasturel}},
  \bibinfo{journal}{Phys. Rev. Lett.} \textbf{\bibinfo{volume}{91}},
  \bibinfo{pages}{195501} (\bibinfo{year}{2003}).

\bibitem[{\citenamefont{Lee et~al.}(2004)\citenamefont{Lee, Gangopadhyay,
  Kelton, Hyers, Rathz, Rogers, and Robinson}}]{KeltonNi}
\bibinfo{author}{\bibfnamefont{G.~W.} \bibnamefont{Lee}},
  \bibinfo{author}{\bibfnamefont{A.~K.} \bibnamefont{Gangopadhyay}},
  \bibinfo{author}{\bibfnamefont{K.~F.} \bibnamefont{Kelton}},
  \bibinfo{author}{\bibfnamefont{R.~W.} \bibnamefont{Hyers}},
  \bibinfo{author}{\bibfnamefont{T.~J.} \bibnamefont{Rathz}},
  \bibinfo{author}{\bibfnamefont{J.~R.} \bibnamefont{Rogers}},
  \bibnamefont{and} \bibinfo{author}{\bibfnamefont{D.~S.}
  \bibnamefont{Robinson}}, \bibinfo{journal}{Phys. Rev. Lett.}
  \textbf{\bibinfo{volume}{93}}, \bibinfo{pages}{037802}
  (\bibinfo{year}{2004}).

\bibitem[{\citenamefont{Schenk et~al.}(2002)\citenamefont{Schenk,
  Holland-Moritz, Simonet, Bellissent, and Herlach}}]{Holland2}
\bibinfo{author}{\bibfnamefont{T.}~\bibnamefont{Schenk}},
  \bibinfo{author}{\bibfnamefont{D.}~\bibnamefont{Holland-Moritz}},
  \bibinfo{author}{\bibfnamefont{V.}~\bibnamefont{Simonet}},
  \bibinfo{author}{\bibfnamefont{R.}~\bibnamefont{Bellissent}},
  \bibnamefont{and} \bibinfo{author}{\bibfnamefont{D.~M.}
  \bibnamefont{Herlach}}, \bibinfo{journal}{Phys. Rev. Lett.}
  \textbf{\bibinfo{volume}{89}}, \bibinfo{pages}{075507}
  (\bibinfo{year}{2002}).

\bibitem[{\citenamefont{Nelson and Widom}(1984)}]{NelsonWidom}
\bibinfo{author}{\bibfnamefont{D.~R.} \bibnamefont{Nelson}} \bibnamefont{and}
  \bibinfo{author}{\bibfnamefont{M.}~\bibnamefont{Widom}},
  \bibinfo{journal}{Nucl. Phys. B} \textbf{\bibinfo{volume}{240}},
  \bibinfo{pages}{113} (\bibinfo{year}{1984}).

\bibitem[{\citenamefont{Sachdev and Nelson}(1984)}]{Sachdev}
\bibinfo{author}{\bibfnamefont{S.}~\bibnamefont{Sachdev}} \bibnamefont{and}
  \bibinfo{author}{\bibfnamefont{D.~R.} \bibnamefont{Nelson}},
  \bibinfo{journal}{Phys. Rev. Lett.} \textbf{\bibinfo{volume}{53}},
  \bibinfo{pages}{1947} (\bibinfo{year}{1984}).

\bibitem[{\citenamefont{DiCicco et~al.}(2003)\citenamefont{DiCicco, Trapananti,
  Faggioni, and Filipponi}}]{Cicco}
\bibinfo{author}{\bibfnamefont{A.}~\bibnamefont{DiCicco}},
  \bibinfo{author}{\bibfnamefont{A.}~\bibnamefont{Trapananti}},
  \bibinfo{author}{\bibfnamefont{S.}~\bibnamefont{Faggioni}}, \bibnamefont{and}
  \bibinfo{author}{\bibfnamefont{A.}~\bibnamefont{Filipponi}},
  \bibinfo{journal}{Phys. Rev. Lett.} \textbf{\bibinfo{volume}{91}},
  \bibinfo{pages}{135505} (\bibinfo{year}{2003}).

\bibitem[{\citenamefont{McGreevy}(2001)}]{Greevy1}
\bibinfo{author}{\bibfnamefont{R.~L.} \bibnamefont{McGreevy}},
  \bibinfo{journal}{J. Phys. Condens. Matter} \textbf{\bibinfo{volume}{13}},
  \bibinfo{pages}{R877} (\bibinfo{year}{2001}).

\bibitem[{\citenamefont{Gurman and McGreevy}(1990)}]{Greevy2}
\bibinfo{author}{\bibfnamefont{S.~J.} \bibnamefont{Gurman}} \bibnamefont{and}
  \bibinfo{author}{\bibfnamefont{R.~L.} \bibnamefont{McGreevy}},
  \bibinfo{journal}{J. Phys. Condens. Matter} \textbf{\bibinfo{volume}{2}},
  \bibinfo{pages}{9463} (\bibinfo{year}{1990}).

\bibitem[{\citenamefont{Wang et~al.}(1997)\citenamefont{Wang, Lu, and
  Li}}]{Wang}
\bibinfo{author}{\bibfnamefont{Y.}~\bibnamefont{Wang}},
  \bibinfo{author}{\bibfnamefont{K.}~\bibnamefont{Lu}}, \bibnamefont{and}
  \bibinfo{author}{\bibfnamefont{C.}~\bibnamefont{Li}}, \bibinfo{journal}{Phys.
  Rev. Lett.} \textbf{\bibinfo{volume}{79}}, \bibinfo{pages}{3664}
  (\bibinfo{year}{1997}).

\bibitem[{\citenamefont{Kresse and Hafner}(1993{\natexlab{b}})}]{VASP}
\bibinfo{author}{\bibfnamefont{G.}~\bibnamefont{Kresse}} \bibnamefont{and}
  \bibinfo{author}{\bibfnamefont{J.}~\bibnamefont{Hafner}},
  \bibinfo{journal}{Phys.\ Rev. B} \textbf{\bibinfo{volume}{47}},
  \bibinfo{pages}{R558} (\bibinfo{year}{1993}{\natexlab{b}}).

\bibitem[{\citenamefont{Kresse and Furthmuller}(1996)}]{VASP2}
\bibinfo{author}{\bibfnamefont{G.}~\bibnamefont{Kresse}} \bibnamefont{and}
  \bibinfo{author}{\bibfnamefont{J.}~\bibnamefont{Furthmuller}},
  \bibinfo{journal}{Phys. Rev. B} \textbf{\bibinfo{volume}{54}},
  \bibinfo{pages}{11169} (\bibinfo{year}{1996}).

\bibitem[{\citenamefont{Torquato et~al.}(2000)\citenamefont{Torquato, Truskett,
  and Debenedetti}}]{MRJ}
\bibinfo{author}{\bibfnamefont{S.}~\bibnamefont{Torquato}},
  \bibinfo{author}{\bibfnamefont{T.~M.} \bibnamefont{Truskett}},
  \bibnamefont{and} \bibinfo{author}{\bibfnamefont{P.~G.}
  \bibnamefont{Debenedetti}}, \bibinfo{journal}{Phys. Rev. Lett.}
  \textbf{\bibinfo{volume}{84}}, \bibinfo{pages}{2064} (\bibinfo{year}{2000}).

\bibitem[{\citenamefont{Blochl}(1994)}]{PAW}
\bibinfo{author}{\bibfnamefont{P.~E.} \bibnamefont{Blochl}},
  \bibinfo{journal}{Phys. Rev. B} \textbf{\bibinfo{volume}{50}},
  \bibinfo{pages}{17953} (\bibinfo{year}{1994}).

\bibitem[{\citenamefont{Kresse and Joubert}(1999)}]{KJ_PAW}
\bibinfo{author}{\bibfnamefont{G.}~\bibnamefont{Kresse}} \bibnamefont{and}
  \bibinfo{author}{\bibfnamefont{D.}~\bibnamefont{Joubert}},
  \bibinfo{journal}{Phys. Rev. B} \textbf{\bibinfo{volume}{59}},
  \bibinfo{pages}{1758} (\bibinfo{year}{1999}).

\bibitem[{\citenamefont{Nose}(1984)}]{Nose}
\bibinfo{author}{\bibfnamefont{S.}~\bibnamefont{Nose}}, \bibinfo{journal}{J.
  Chem. Phys} \textbf{\bibinfo{volume}{81}}, \bibinfo{pages}{511}
  (\bibinfo{year}{1984}).

\bibitem[{\citenamefont{Waseda}(1980)}]{Waseda}
\bibinfo{author}{\bibfnamefont{Y.}~\bibnamefont{Waseda}}, \bibinfo{journal}{The
  Structure of Non-Crystalline Materials}  (\bibinfo{year}{1980}),
  \bibinfo{note}{(McGraw-Hill, New York)}.

\bibitem[{\citenamefont{Stillinger and Weber}(1982)}]{Stillinger}
\bibinfo{author}{\bibfnamefont{F.~H.} \bibnamefont{Stillinger}}
  \bibnamefont{and} \bibinfo{author}{\bibfnamefont{T.~A.} \bibnamefont{Weber}},
  \bibinfo{journal}{Phys. Rev. A} \textbf{\bibinfo{volume}{25}},
  \bibinfo{pages}{978} (\bibinfo{year}{1982}).

\bibitem[{\citenamefont{Eder et~al.}(1990)\citenamefont{Eder, Erdpresser,
  Kunsch, Stiller, and Suda}}]{Eder}
\bibinfo{author}{\bibfnamefont{O.~J.} \bibnamefont{Eder}},
  \bibinfo{author}{\bibfnamefont{E.}~\bibnamefont{Erdpresser}},
  \bibinfo{author}{\bibfnamefont{B.}~\bibnamefont{Kunsch}},
  \bibinfo{author}{\bibfnamefont{H.}~\bibnamefont{Stiller}}, \bibnamefont{and}
  \bibinfo{author}{\bibfnamefont{M.}~\bibnamefont{Suda}}, \bibinfo{journal}{J.
  Phys. F} \textbf{\bibinfo{volume}{10}}, \bibinfo{pages}{183}
  (\bibinfo{year}{1990}).

\bibitem[{\citenamefont{Baxter}(1970)}]{Baxter}
\bibinfo{author}{\bibfnamefont{R.~J.} \bibnamefont{Baxter}},
  \bibinfo{journal}{J. Chem. Phys} \textbf{\bibinfo{volume}{52}},
  \bibinfo{pages}{4559} (\bibinfo{year}{1970}).

\bibitem[{\citenamefont{Jolly et~al.}(1976)\citenamefont{Jolly, Freasier, and
  Bearman}}]{Jolly}
\bibinfo{author}{\bibfnamefont{D.~J.} \bibnamefont{Jolly}},
  \bibinfo{author}{\bibfnamefont{B.~C.} \bibnamefont{Freasier}},
  \bibnamefont{and} \bibinfo{author}{\bibfnamefont{R.~J.}
  \bibnamefont{Bearman}}, \bibinfo{journal}{Chem. Phys}
  \textbf{\bibinfo{volume}{15}}, \bibinfo{pages}{237} (\bibinfo{year}{1976}).

\bibitem[{\citenamefont{Foiles and Ashcroft}(1984)}]{Ashcroft}
\bibinfo{author}{\bibfnamefont{S.~M.} \bibnamefont{Foiles}} \bibnamefont{and}
  \bibinfo{author}{\bibfnamefont{N.~W.} \bibnamefont{Ashcroft}},
  \bibinfo{journal}{J. Chem. Phys} \textbf{\bibinfo{volume}{81}},
  \bibinfo{pages}{6140} (\bibinfo{year}{1984}).

\bibitem[{\citenamefont{Luo et~al.}(2004)\citenamefont{Luo, Sheng, Alamgir,
  Bai, He, and Ma}}]{Ni-Ag}
\bibinfo{author}{\bibfnamefont{W.~K.} \bibnamefont{Luo}},
  \bibinfo{author}{\bibfnamefont{H.~W.} \bibnamefont{Sheng}},
  \bibinfo{author}{\bibfnamefont{F.~M.} \bibnamefont{Alamgir}},
  \bibinfo{author}{\bibfnamefont{J.~M.} \bibnamefont{Bai}},
  \bibinfo{author}{\bibfnamefont{J.~H.} \bibnamefont{He}}, \bibnamefont{and}
  \bibinfo{author}{\bibfnamefont{E.}~\bibnamefont{Ma}}, \bibinfo{journal}{Phys.
  Rev. Lett.} \textbf{\bibinfo{volume}{92}}, \bibinfo{pages}{145502}
  (\bibinfo{year}{2004}).

\bibitem[{\citenamefont{Clarke and Jonsson}(1993)}]{Clarke}
\bibinfo{author}{\bibfnamefont{A.~S.} \bibnamefont{Clarke}} \bibnamefont{and}
  \bibinfo{author}{\bibfnamefont{H.}~\bibnamefont{Jonsson}},
  \bibinfo{journal}{Phys. Rev. E} \textbf{\bibinfo{volume}{47}},
  \bibinfo{pages}{3975} (\bibinfo{year}{1993}).

\bibitem[{\citenamefont{Jakse}(2004)}]{Jakse_alloy1}
\bibinfo{author}{\bibfnamefont{N.}~\bibnamefont{Jakse}},
  \bibinfo{journal}{Phys. Rev. Lett.} \textbf{\bibinfo{volume}{93}},
  \bibinfo{pages}{207801} (\bibinfo{year}{2004}).

\bibitem[{\citenamefont{Jakse et~al.}(2005)\citenamefont{Jakse, Bacq, and
  Pasturel}}]{Jakse_alloy2}
\bibinfo{author}{\bibfnamefont{N.}~\bibnamefont{Jakse}},
  \bibinfo{author}{\bibfnamefont{O.~L.} \bibnamefont{Bacq}}, \bibnamefont{and}
  \bibinfo{author}{\bibfnamefont{A.}~\bibnamefont{Pasturel}},
  \bibinfo{journal}{J. Chem. Phys.} \textbf{\bibinfo{volume}{123}},
  \bibinfo{pages}{104508} (\bibinfo{year}{2005}).

\bibitem[{\citenamefont{Sheng et~al.}(2006)\citenamefont{Sheng, Luo, Alamgir,
  Bai, and Ma}}]{MaNature}
\bibinfo{author}{\bibfnamefont{H.~W.} \bibnamefont{Sheng}},
  \bibinfo{author}{\bibfnamefont{W.~K.} \bibnamefont{Luo}},
  \bibinfo{author}{\bibfnamefont{F.~M.} \bibnamefont{Alamgir}},
  \bibinfo{author}{\bibfnamefont{J.~M.} \bibnamefont{Bai}}, \bibnamefont{and}
  \bibinfo{author}{\bibfnamefont{E.}~\bibnamefont{Ma}},
  \bibinfo{journal}{Nature} \textbf{\bibinfo{volume}{439}},
  \bibinfo{pages}{419} (\bibinfo{year}{2006}).

\bibitem[{\citenamefont{Ganesh and Widom}(2006)}]{Fe_alloy}
\bibinfo{author}{\bibfnamefont{P.}~\bibnamefont{Ganesh}} \bibnamefont{and}
  \bibinfo{author}{\bibfnamefont{M.}~\bibnamefont{Widom}},
  \bibinfo{journal}{preprint}  (\bibinfo{year}{2006}).

\end{thebibliography}

\end{document}